\documentclass[12pt ]{article}
\usepackage{amsmath}
\usepackage{graphicx}
\usepackage{hyperref}
\usepackage{lmodern}
\usepackage{amssymb}
\usepackage{booktabs}
\usepackage{tikz}
\usetikzlibrary{patterns}
\usepackage{pdfpages}

\newcommand{\be}{\begin{equation}}
\newcommand{\ee}{\end{equation}}

\newcommand{\beqa}{\begin{eqnarray}}
\newcommand{\eeqa}{\end{eqnarray}}
\newcommand{\nn}{\nonumber}

\def\CE {{\cal E}}
\def\CF {{\cal F}}
\def\CG {{\cal G}}
\def\CH {{\cal H}}

\def\CL {{\cal L}}

\begin{document}

\newpage

\setlength{\baselineskip}{7mm}
\begin{titlepage}
 
\begin{flushright} 
 {\tt NRCPS-HE-52-2024} 
\end{flushright}

\begin{center}
{\Large ~\\{\it    Landscape of QCD Vacuum 
}

}

\vspace{3cm}

{\sl George Savvidy

\centerline{${}$ \sl Institute of Nuclear and Particle Physics}
\centerline{${}$ \sl Demokritos National Research Center, Ag. Paraskevi,  Athens, Greece}

}
 
\end{center}
\vspace{3cm}

\centerline{{\bf Abstract}}

We found new solutions of the sourceless Yang-Mills equation describing the superposition of chromomagnetic vortices of oppositely oriented magnetic fluxes. These gauge field configurations have constant energy densities and are separated by potential barriers forming a complicated landscape.   It is suggested that the solutions describe the condensate of chromomagnetic  vortices and represent a dual analog of the Cooper pairs condensate in a superconductor. In the presence of an Abelian field and in a particular limit the solutions reduce to flat connections of zero energy density and are forming a complicated potential landscape of the QCD vacuum.   A possible tunnelling  transition between  these superfluxon  flat configurations and the flat configurations with non-vanishing  Chern-Pontryagin index will wash out the CP violating $\theta$ angle to zero, dynamically restoring CP symmetry.

  \vspace{12pt}

\noindent

\end{titlepage}

\pagestyle{plain}

\section{\it  Magnetic monopoles and magnetic vortices }

It is largely expected that the confining QCD vacuum should represent a dual superconductor state with a condensate of the monopole-antimonopole pairs in analogy with Cooper pairs in ordinary superconductors  \cite{tHooft:1974kcl, Polyakov:1974ek, tHooft:1981bkw, tHooft:1979rtg, Mandelstam:1978ed, tHooft1977,  Mandelstam:1980ii,cmp/1104178138, https://doi.org/10.1112/plms/s3-55.1.59, Gukov:2006jk, Gukov:2008sn, Kapustin:2005py, Kapustin:2006pk, DeBenedictis:2024dts}.  The pure Yang Mills theory does not have adjoint scalar fields and therefore does not admit explicit 't Hooft-Polyakov monopole solutions \cite{tHooft:1974kcl, Polyakov:1974ek} as well as  Nielsen-Olesen magnetic vortices \cite{Nielsen:1973cs}.  Whether or not there is a  condensation of the monopole-antimonopole pairs or of the magnetic vortics depends on the details of the dynamics, on the  existence of solutions with long-range topological structure in the pure Yang Mills theory.  The dynamics of magnetic fluxes  must be an important ingredient of the confinement mechanism in addition to the  topological arguments.  The question is if such long-range topological structure exists in the pure Yang Mille theory,  in QCD.  

The guideline in the search of the appropriate solutions can be found by analising   the topological structure  of the  't Hooft-Polyakov monopole  \cite{tHooft:1974kcl, Polyakov:1974ek} and Nielsen-Olesen  magnetic vortex solutions  \cite{Nielsen:1973cs}.  The existence of 't Hooft-Polyakov monopole solution   in spontaneously broken $SU(2)$ gauge field theory  is due to the presence of adjoint  scalar field $\phi_a, a=1,2,3$ and of the associated singularities.  The electromagnetic field strength was defined by 't Hooft as \cite{tHooft:1974kcl}
\be\label{abelean}
G_{\mu\nu}=   n^a G^a_{\mu\nu}
+ {1\over g }\epsilon^{abc} n^a   \nabla_{\mu}  n^b \nabla_{\nu} n^c  \equiv \partial_{\mu} B_{\nu} - \partial_{\nu} B_{\mu} + {1\over g }\epsilon^{abc} n^a   \partial_{\mu}  n^b  \partial_{\nu}  n^c,~~~~~ n^a = { \phi^a \over \vert \phi \vert},
\ee 
where $\nabla_{\mu}  n^a= \partial_{\mu} n^a -g \epsilon^{abc} A^b_{\mu} n^c$,  $B_{\mu}= A^{a}_{\mu} n^a$ and $n^a$ is a unit colour vector.  The definition (\ref{abelean}) satisfies the Maxwell equations, except for the spacetime point where the scalar field vanishes, $\phi_a(x) = 0$ , and  the field $n^a(x)$ develops a singularity. {\it This    indicates the existence and the location of a monopole in this theory} \cite{tHooft:1974kcl, Polyakov:1974ek}.  This   follows from the topologically  conserved  current\footnote{The first term in (\ref{abelean}) does not contribute to the topological current if there are no Dirac string-like singularities in $A_{\mu}$. }
\be\label{topdensity}
K_{\mu}={1\over 2 }  \epsilon_{\mu\nu\lambda\rho}  \partial_{\nu}  G_{\lambda\rho}= {1\over 2 g }  \epsilon_{\mu\nu\lambda\rho} \epsilon^{abc}  \partial_{\nu}  n^a  \partial_{\lambda} n^b \partial_{\rho} n^c~,~~~~~~ \partial_{\mu} K_{\mu}=0.
\ee
The singularity appears at the point where the scalar field  $\phi^a= n^a u(r)$   
vanishes,  $u(0) =  0$,   
$
 n^a(x) ={x^a \over r} 
$
and   $K_0={4 \pi \over g} \delta^3(\vec{x})$.  The topological charge  is proportional  to the winding number
\be\label{topdeg}
g_m = \int_{R^3} d^3x K_0 ={1 \over 2 g}   \int_{S^2 } d^2 \sigma_i \epsilon_{ijk} \epsilon^{abc}     n^a  \partial_{j} n^b \partial_{k} n^c  = {4\pi \over g} 
\ee 
and it is also equal  to the magnetic flux of a single monopole solution $ A^a_i \underset{r \rightarrow \infty}\rightarrow  - \epsilon^{aij} x^j / g r^2  $:
 \be\label{magflax}
 H_i = {x_i \over g r^3}, ~~~~~~g_m= \int H_i dS_i = {4\pi \over g}.
 \ee
 The monopole charge $g_m$ is characterised  by a topological degree (\ref{topdeg}) and by a  total magnetic flux   (\ref{magflax}).  A similar consideration is valid for the   Nielsen-Olesen  magnetic vortex solution  in the Abelian-Higgs model \cite{Nielsen:1973cs, Zumino:1974beb, Nambu:1974zg}.  {\it The magnetic monopoles and magnetic vortices emerge as physical objects exhibiting  themselves through the  zeros of the scalar field and  the associated singularities} \cite{tHooft:1974kcl}.   The lesson that follows from the above discussion is that one can trace the existence of new magnetic structures in the pure Yang-Mills theory by investigating possible singularities of the gauge fields \cite{  tHooft:1981bkw, tHooft:1979rtg, Mandelstam:1978ed, tHooft1977,  Mandelstam:1980ii,cmp/1104178138, https://doi.org/10.1112/plms/s3-55.1.59, Gukov:2006jk, Gukov:2008sn, Kapustin:2005py, Kapustin:2006pk}. 
 
 We shall demonstrate that in the pure Yang Mills theory  the field strength tensor can have the same structure  as in the Yang-Mills-Higgs model (\ref{abelean}) but now the role of the unit colour vector field $n^a(x)$ is not connected with any scalar field but instead with the Yang Mills field itself.
 The nontrivial topological field configurations of the gauge field and their singularities can be found by considering the solutions of the {\it covariantly constant field equation} \cite{Batalin:1976uv, Savvidy:1977as, Matinyan:1976mp, Brown:1975bc, Duff:1975ue,Savvidy:2024sv, Savvidy:2024ppd}. The exact  solutions of the sourceless Yang Mills equation have nontrivial topological structure  and singularities that are distributed over two-dimensional sheets and cylinders representing a lattice of superposed  magnetic vortices  \cite{Savvidy:2024sv, Savvidy:2024ppd}.  The location of field singularities  is invariant with respect to the continuous gauge transformations and characterises the moduli space of the solutions. 
  
   The solutions represent non-perturbative chromomagnetic vortices, the flux tubes similar in their structure to superposed Nielsen-Olesen magnetic vortices \cite{Nielsen:1973cs} uniformly distributed over the whole space \cite{Savvidy:2024sv, Savvidy:2024ppd}.  These gauge field configurations are stretched along the potential valleys of a constant energy density and are separated by potential barriers  forming a complicated potential landscape. It is suggested that the solutions describe the condensate of chromomagnetic  vortices  of oppositely oriented magnetic fluxes  representing a dual analog of the Cooper pairs condensate in a superconductor. This consideration leads to a description of the vacuum state of the Yang-Mills theory as having a richer topological structure than previously thought.

\section{\it General solution of covariantly constant gauge field equation }

The covariantly constant gauge fields are defined by the equation \cite{Batalin:1976uv, Savvidy:1977as, Matinyan:1976mp, Brown:1975bc, Duff:1975ue}  
\be\label{YMeqcov}
\nabla^{ab}_{\rho} G^{b}_{\mu\nu} =0,
\ee
where 
$
G^{a}_{\mu\nu} =  \partial_{\mu} A^{a}_{\nu} - \partial_{\nu} A^{a}_{\mu}
 - g \varepsilon^{abc} A^{b}_{\mu} A^{c}_{\nu},
$ 
$
\nabla^{ab}_{\mu}(A)=
  \delta^{ab} \partial_{\mu}   - g \varepsilon^{acb} A^{c}_{\mu}
$
and are the solutions of the sourceless  Yang-Mills equation $\nabla^{ab}_{\mu} G^{b}_{\mu\nu} =0$ as well\footnote{ The effective Lagrangian is gauge invariant only on sourceless-vacuum fields \cite{Batalin:1976uv,Batalin:1979jh}.}.  Here we will consider the $SU(2)$ algebra, the consideration can be extended to other algebras as well. By taking the covariant derivative $\nabla^{ca}_{\lambda}$ of the l.h.s (\ref{YMeqcov}) and interchanging the derivatives one can get
$
 [G_{\lambda\rho},  G_{\mu\nu}] =0,
$
which means that the field strength tensor factorises into the product  of Lorentz tensor and colour unit vector in the direction of the Cartan's sub-algebra:
\be\label{covconfac}
G^{a}_{\mu\nu}(x)= G_{\mu\nu}(x) n^a(x). 
\ee
 Both fields can depend on the space-time coordinates. The well known solution of (\ref{YMeqcov}) has the following form \cite{Batalin:1976uv, Savvidy:1977as, Matinyan:1976mp, Brown:1975bc, Duff:1975ue}: 
 \be\label{consfield}
A^{a}_{\mu} = - {1\over 2} F_{\mu\nu} x_{\nu}   n^a , 
\ee
where $F_{\mu\nu} $ and $n^a$ are space-time independent parameters, $ n^{a} n^{a} =1$. It is convenient to call this solution  as  a "constant Abelian chromomagnetic field" \footnote{The  solution has six parameters $F_{\mu\nu}$, four translations $x_{\nu} \rightarrow x_{\nu} + x_{0 \nu} $ and two parameters $n^a$ in the case of $SU(2)$ group.} because $ n^a$ is a constant colour vector.  The general solutions of the equation (\ref{YMeqcov}) were found recently in  \cite{Savvidy:2024sv, Savvidy:2024ppd}.     The new solutions can be obtained through the nontrivial space-time dependence of the unit vector $n^a(x)$.  Considering the   Ansatz  \cite{ tHooft:1974kcl,Cho:1979nv, Cho:1980nx, Cho:2010zzb, Corrigan:1975zxj, Biran:1987ae, Savvidy:2024sv, Savvidy:2024ppd}
\be\label{choansatz}
A^{a}_{\mu} =  B_{\mu} n^{a}  +
{1\over g} \varepsilon^{abc} n^{b} \partial_{\mu}n^{c},
\ee
where $B_{\mu}(x)$  is the Abelian Lorentz vector and $n^{a}(x)$ is a space-time dependent colour unit vector
$
n^{a} n^{a} =1,
$
$n^{a} \partial_{\mu} n^{a} =0,
$
one can observe that the field strength tensor factorises  \cite{Cho:1979nv, Cho:1980nx}:
\be\label{chofact}
G^{a}_{\mu\nu} =  ( F_{\mu\nu} + {1\over g} S_{\mu\nu})~ n^{a} \equiv G_{\mu\nu}(x)~ n^a(x),
\ee
where
\be\label{spacetimefields}
F_{\mu\nu}= \partial_{\mu} B_{\nu} - \partial_{\nu} B_{\mu},~~~~~~~~~
S_{\mu\nu}= \varepsilon^{abc} n^{a} \partial_{\mu} n^{b} \partial_{\nu} n^{c}.
\ee
This form of the field strength tensor $G_{\mu\nu}(x)$ is identical to the 't Hooft form of the field strength tensor (\ref{abelean}) as well as  to the factorisation form (\ref{covconfac}) of the covariantly constant field strength tensor.  It is therefore natural to search  solutions of (\ref{YMeqcov}) in the form (\ref{choansatz}).   In that case (\ref{YMeqcov})  reduces to the following equation:
 \be
   \partial_{\rho}  ( F_{\mu\nu} + {1\over g} S_{\mu\nu})=0,
 \ee
 meaning that the sum of the terms in the brackets should be a constant tensor:
$
G_{\mu\nu}=  F_{\mu\nu} + {1\over g} S_{\mu\nu}.  
$
It is useful to parametrise the unit vector in terms of spherical angles:
\be\label{unitvector}
n^a = (\sin\theta \cos\phi, \sin\theta \sin\phi, \cos\theta ),
\ee
and express $S_{\mu\nu}$ in terms of spherical angles as well
$
S_{\mu\nu} = \sin\theta ( \partial_{\mu}  \theta  \partial_{\nu} \phi  - \partial_{\nu}  \theta   \partial_{\mu} \phi).
$ 
Let us consider  the solutions that have constant space components $S_{ij}$  and  $F_{ij}$ with time components  $S_{0i}$ and $F_{0i}$ equal to zero. These solutions represent the pure chromomagnetic fields, and the equation (\ref{YMeqcov}) reduces to the following system of partial differential equations:
\beqa\label{gen}
S_{12}= \sin \theta (\partial_1 \theta \partial_2 \phi - \partial_2 \theta \partial_1 \phi ), \nn\\
S_{23}= \sin \theta (\partial_2 \theta \partial_3 \phi - \partial_3 \theta \partial_2 \phi ), \nn\\
S_{13}= \sin \theta (\partial_1 \theta \partial_3 \phi - \partial_3 \theta \partial_1 \phi ).
\eeqa
 The linear combination of these equations defines the  angle $\phi$ as an arbitrary function of  the variable 
$Y=  b_1 x +b_2 y + b_3 z -b_0 t$,
thus
$
 \phi(Y)  =\phi( b\cdot x ),
$
where $b_{\mu}, \mu=0,1,2,3$ are arbitrary real numbers. After substituting the above function into the equations (\ref{gen}) one can observe that the angle variable $\theta$ is a function of the alternative variable $X= a\cdot x $, thus
$
 \theta(X)=  \theta( a\cdot x ),
$
where $a_{\mu}, \mu=0,1,2,3$ are arbitrary real numbers as well. It follows that  the equations (\ref{gen})  reduce to the following differential  equations:
\be\label{genecovfie}
S_{ij} = a_i \wedge b_j  \sin\theta(X) ~ \theta(X)^{'}_X   ~ \phi(Y)^{'}_Y , 
\ee
where the derivatives are over the respective arguments.  The solutions with a constant tensor $S_{ij}$  should fulfil the following equation:
\be\label{ansatz7}
\sin\theta(X) ~ \theta(X)^{'}_X   ~ \phi(Y)^{'}_Y =1,
\ee
so that 
$
S_{ij} = a_i \wedge b_j 
$
and 
\be\label{genenergden}
G_{ij}= F_{ij} + {1\over g}  a_i \wedge b_j,~~~~~~\epsilon =  {1\over 4 }G^{a}_{ij} G^{a}_{ij}  =   {(g  \vec{H} -   \vec{a} \times \vec{b} )^2 \over 2 g^2}.
\ee
The minus sign takes place when three vectors  $ (\vec{H}, \vec{a}, \vec{b})$ are forming an orthogonal right-oriented frame and plus sign for left-oriented frame \cite{Savvidy:2024sv, Savvidy:2024ppd}.
The variables in (\ref{genecovfie})  are independent, therefore we can choose an arbitrary function  $\theta$ and define the function $\phi$ by integration. Let the $\theta(X) $ be an arbitrary function of $X$, then $\phi = Y/\sin \theta(X) \theta(X)^{'}_X$, and we have the following general solution for the colour unit vector (\ref{unitvector}):
\be\label{generasol}
n^a(\vec{x})= \{\sin \theta(X)  \cos\Big({Y \over \theta(X)^{'} \sin \theta(X)} \Big),~\sin \theta(X)  \sin\Big({Y \over \theta(X)^{'}  \sin \theta(X) }\Big),~ \cos \theta(X)   \}.
\ee 
The explicit form of the vector potential $A^a_{\mu}$ can be obtained by substituting the  unit colour vector (\ref{generasol}) into the (\ref{choansatz}).  The arbitrary function $\theta(X)$ in the equation (\ref{generasol}) defines the moduli space of the solutions.  The singularities are located on the planes $X_s$, where the  $\sin\theta(X)$  or $\theta(X_s)^{'}$ vanishe:  \be\label{singplanes} 
\theta(X_s) = 2 \pi N, ~~~N=0,\pm 1, \pm 2...., ~~~\text{or}~~~\theta(X_s)^{'} = 0
\ee
and  $\cos/\sin\Big({Y \over \theta(X)^{'} \sin\theta(X)}\Big)$ are fast oscillating trigonometric functions\footnote{There is some analogy with the Schwarzschild solution in gravity, where the solution is asymptotically flat and regular at infinity while it  has metric singularity at the Schwarzschild radius $ r_g=2MG/c^2$ of the event horizon. This metric singularity is not a physical one because the Riemann curvature tensor is regular at $r=r_g$. Here as well, we have singularities of the gauge field and of the field strength tensor while the energy density is a regular function.}. Our aim is to describe the  moduli space of the covariantly constant gauge fields  defined by the equations (\ref{YMeqcov}), (\ref{choansatz})  and (\ref{generasol}) and investigate their physical properties.

\begin{figure}
 \centering
\includegraphics[angle=0,width=8cm]{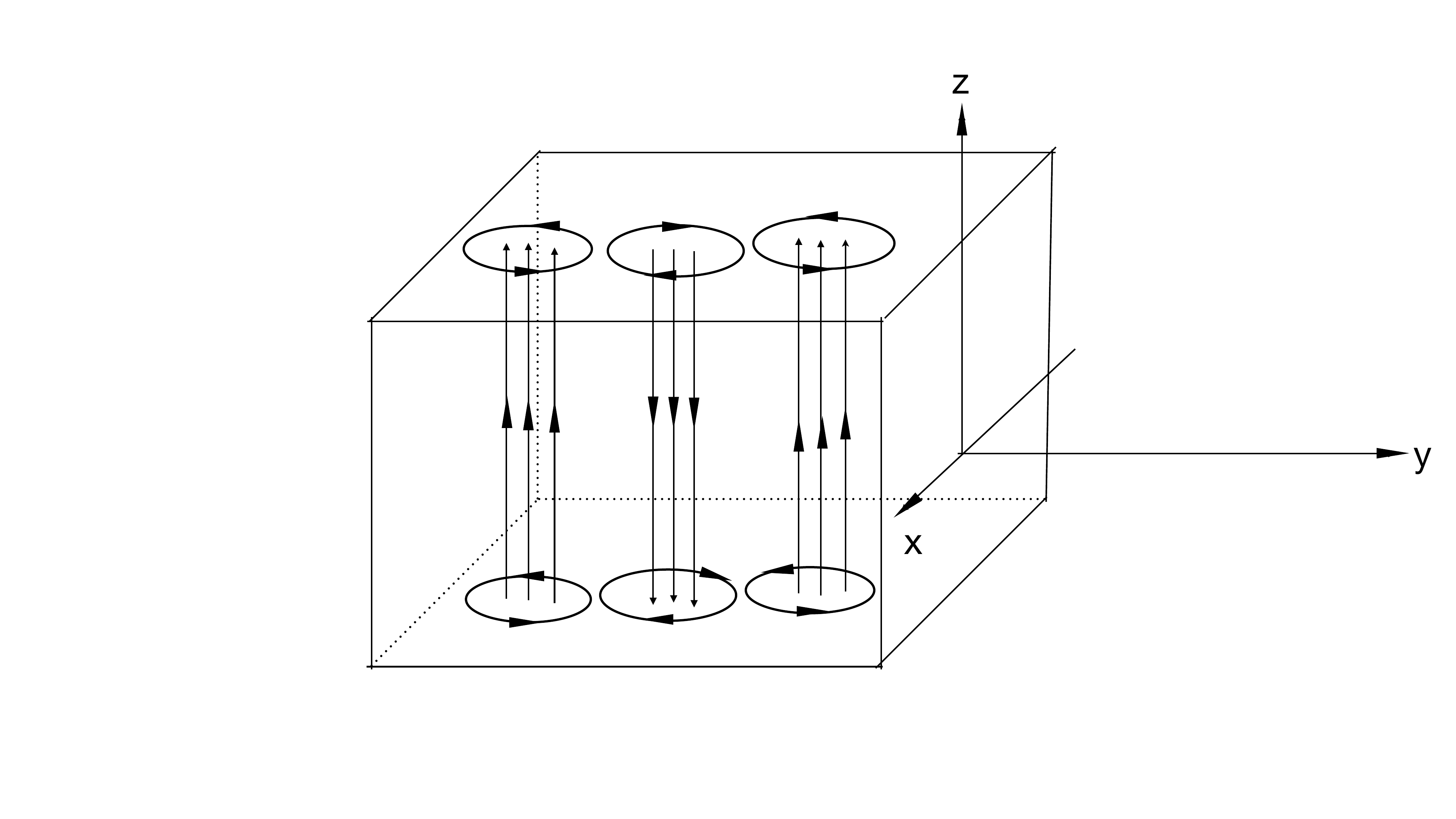}
\centering
\caption{The figure demonstrates a finite part of an infinite sheet of finite thickness ${2\over a}$ in the direction of the $x$ axis of the solution (\ref{magsheet}). It is  filled by  parallel chromomagnetic flux tubes. Each tube of the square area ${2\over a}  {  \pi \over b}$ carries the magnetic flux ${4 \pi \over g}$.  The circuits show the flow of the conserved current   $J^a_{k}=g \epsilon^{abc} A^b_{j} G^c_{ik} $ and the arrows show the flow of the vorticity $\omega^a_i = \epsilon_{ijk} \partial_j J^a_k $.    }
\label{fig1} 
\end{figure}
 The general solution (\ref{generasol}) for the vector potential $A^a_{\mu}$   (\ref{choansatz})  depends on two coordinates $X$ and $Y$. There are two subclasses of physically interesting solutions: the time independent solutions $a_0 =b_0= 0$ describing stationary magnetic fluxes distributed in 3d-space and time-dependent solutions $a_0 \neq 0$,  $b_0 \neq 0$ describing propagation of strings and branes.   For the sake of transparency and compactness of the subsequent formulas we will identify this plane with the   $(x,y)$ plane. Thus we are considering the vectors  $a_{\mu}=(0,a,0,0)$ and $b_{\nu}=(0,0,b,0)$, so that  $\theta(x) = f(a x)$, $\phi(x,y) = b y /f^{'}(ax) \sin f(ax)   $.  The gauge field (\ref{choansatz}) together with the Abelian field $B_{1}=H y$ will take the following form:
\beqa\label{magneticsheetsolution1}
A^{a}_{i}(x,y) &=&   {1\over g} \left\{
\begin{array}{ccccc}   
a \Big( b y ( ({g H \over a b}-1)  \sin f    +{1 \over \sin f  } )     \cos ({b y  \over f^{'}  \sin f }) - f^{'}  \sin ({b y  \over f^{'} \sin f }) +  b y  {  f^{''}   \over f^{'2}} \cos f  \cos ({b y  \over f^{'}  \sin f }), \\ 
~~ b y ( ({g H \over a b}-1)  \sin f    +{1 \over \sin f  } )      \sin({b y  \over f^{'}  \sin f }) + f^{'}   \cos({b y  \over f^{'}  \sin f }) + b y {  f^{''}   \over f^{' 2}} \cos f  \sin ({b y  \over f^{'}  \sin f}), ~~~\\
    b  y (({g H \over a b}-1) \cos f -   { f^{''}  \over f^{'2}}  \sin f  )\Big)  \\ 
{b \over f^{'}} \Big(-\cos f  \cos ({b y  \over f^{'} \sin f}),-\cos f  \sin ({b y  \over f^{'} \sin f}), ~  \sin f \Big)\\
(0,0,0),
\end{array} \right. ,  
\eeqa 
where $i=1,2,3$ and the derivatives are over the whole argument $ax$. The  $A^a_0=0$ and the singularities are at (\ref{singplanes}). One can verify explicitly  that it is a solution of the Yang Mills equation \cite{Savvidy:2024sv, Savvidy:2024ppd}. The nonzero component of the  field strength tensor $G^a_{\mu\nu}$ has the  following form:
 \be\label{consfielstr}
 G^{a}_{12}(x,y) =  {a b - g H \over g}   ~ n^a(x,y),
 \ee
 and the energy density of the chromomagnetic field is a regular function  and is a space-time constant:
 \be\label{energydenscont}
\epsilon =  {1\over 4 }G^{a}_{ij} G^{a}_{ij} =    { (gH-a b )^2 \over 2 g^2}.
 \ee  
There are two important limiting solutions when $H=0$  and $g H = a b$. In the first case we have solutions which have constant energy densities $\epsilon =   (a b )^2 / 2 g^2$,  like the solution (\ref{consfield}), and are separated from each other by potential barriers. In the second case we will get new {\it vacuum solutions},  because for them $G_{\mu\nu} =0$ and $\epsilon =0$. The solutions represent the flat connections characterised by the vectors $(\vec{H}, \vec{a}, \vec{b})$.  They are separated from the $A_{\mu}=0$ vacuum solution also by potential barriers. In the next sections we will analyse the shape of these barriers.

The non-vanishing components of the conserved current $J^a_{\mu} = g \epsilon^{abc} A^b_{\nu} G^c_{\nu\mu}$ at $H=0$ are\footnote{This current is conserved on the solutions of the Yang Mills equation $\nabla^{ab}_{\mu} G^{b}_{\mu\nu} =0$.} 
 \beqa\label{conscurrent}
 J^a_1&=&{a b^2\over g  f^{'} } \Big( \sin({b y \over  f^{'} \sin f}), -\cos({b y \over   f^{'} \sin f}),0 \Big);~\\
J^1_2&=&{a^2 b\over g} \Big( f^{'} \cos f \cos({b y \over f^{'} \sin f}) + b y   \cot f \sin({b y \over f^{'} \sin f}) + b y {f^{''}\over f^{' 2}} \sin({b z \over f^{'} \sin f})   \Big), \nn\\
 J^2_2&=&{a^2 b\over g} \Big(f^{'}  \cos f \sin({b y \over f^{'} \sin f}) -b y \cot f  \cos({b y \over f^{'} \sin f}) - b y {f^{''}\over f^{' 2} } \cos({b y \over f^{'} \sin f})   \Big),~\nn\\
 J^3_2&=&- {a^2 b\over g}  f^{'} \sin f. \nn
 \eeqa
One can check that $ \partial_{\mu} J^a_{\mu}=    \partial_x J^a_1 + \partial_y J^a_2 =0 $ and that the vorticity $\omega^a_i = \epsilon_{ijk} \partial_j J^a_k $ of the current  is nonzero and is  singular at the location of the vortices.   The space-time structure of the solution  (\ref{choansatz}), (\ref{magneticsheetsolution1})  is reminiscent of the "spaghetti"-type configurations found as an approximate solution of the Yang Mills equation in a background field (\ref{consfield}) \cite{  Nielsen:1978rm,  Ambjorn:1978ff, Nielsen:1978tr, Nielsen:1979xu, Ambjorn:1979xi, Ambjorn:1980ms, Olesen:1981zp, Ambjorn:1988tm, Ambjorn:1989bd, Ambjorn:1989sz, Arodz:1980gh}.  The  equations (\ref{choansatz}),  (\ref{consfield}), (\ref{generasol}) and (\ref{magneticsheetsolution1}) represent an {\it exact non-perturbative solution of the YM equation in the background chromomagnetic field (\ref{consfield})} \footnote{An early attempt to find a larger class of  space-homogeneous vacuum  Yang-Mills fields was made in \cite{Baseian:1979zx, Savvidy:1982wx, Savvidy:1982jk, Matinyan:1981ys, Matinyan:1981dj,Banks:1996vh}.  It was shown that space-homogeneous vacuum fields exhibit deterministic  chaos \cite{Savvidy:2020mco}. The vacuum fields were also considered in \cite{ Pak:2020izt, Pak:2020obo, Pak:2017skw, Pak:2020fkt,  Anous:2017mwr, Kim:2016xdn, Milshtein:1983th, Olesen:1981zp,  Apenko:1982tj, Reuter:1994yq, Reuter:1997gx,   Wu:1975vq, Wu:1967vp}. The   approximate solutions  in the background field  (\ref{consfield}) were investigated in \cite{  Nielsen:1978rm,  Ambjorn:1978ff, Nielsen:1978tr, Nielsen:1979xu, Ambjorn:1979xi, Ambjorn:1980ms, Olesen:1981zp, Ambjorn:1988tm, Ambjorn:1989bd, Ambjorn:1989sz, Arodz:1980gh} and have a "spaghetti"-type structure of magnetic flux tubes. }.

 Let us consider solutions through which one can expose the essential properties of the general solution.  To obtain a particular solution in an explicit form we have to choose the function $\theta(X)$. Considering  $\theta(X)=\arcsin(\sqrt{1- (a\cdot x)^2})$ we are obtaining a "chromomagnetic flux sheet"  solution  \cite{Savvidy:2024sv, Savvidy:2024ppd}
\be\label{polsol}
n^a(x)= \{ \sqrt{1-(a\cdot x)^2} \cos(b\cdot x),~ \sqrt{1-(a\cdot x)^2} \sin(b\cdot x),~ (a\cdot x)  \},
\ee 
which represents a {\it non-perturbative  magnetic sheet of a finite thickness $  2/\vert a \vert $}, and the corresponding gauge field (\ref{choansatz}) has the following form:
\beqa\label{magsheet} 
A^{a}_{i}(x,y) = {1\over g} \left\{
\begin{array}{llll}   
 {1 \over \sqrt{1-(a x)^2} }\Big(  a \sin b y - gH y  (1-(a x)^2) \cos b y ,\\
 -   a \cos b y  - gH y  (1-(a x)^2)\sin b y, -g H y a x  \sqrt{1-(a x)^2}\Big)& \\
b \sqrt{1-(a x)^2} \Big(-a   x \cos b y , - a    x \sin b y ,  \sqrt{1-(a x)^2}\Big)&\\
(0,0,0),~~~~~~~~~~~~~~~~~~~~~~~~~~~~~~~~~~~~~~~~~~~~~~~~~~~~~~~~~~~~~~~~~~~~~~~~ (a  x)^2 <  1 &
\end{array} \right.   
\eeqa 
where $\vec{a}=(a,0,0)$, $\vec{b}=(0,b,0)$ and $A^{a}_{\mu} =  0$ when $ (a  x)^2 \geq 1 \nn$. The current vorticity $\omega^a_i = \epsilon_{ijk} \partial_j J^a_k $  
\be
\omega^a_3 ={1\over g}  { (a b - gH)(a^2 +b^2(1- a^2 x^2)^2 ) \over (1-(a x)^2)^{3/2}  }  \Big( \cos b y,  \sin b y,0 \Big),~~~ (a  x)^2 <  1 
\ee
is  singular at the location of the vortices $x=\pm 1/a$.  There is no energy flow from the magnetic sheet in the direction transversal to the sheet  because the Poynting vector vanishes, $\vec{E^a} \times \vec{H^a} =0$.  This solution is similar to the superposition of the Nielsen-Olesen magnetic flux tubes and is supported without presence of any Higgs field (see Fig.\ref{fig1}).  The magnetic flux that is defined by the equation\footnote{The $W(L)$ is a character of the SU(2) representations $\chi_j = {\sin(j+1/2) \Phi  \over \sin(\Phi/2) }$ and for $j=1/2$ is $ \chi_{1/2} = 2\cos(\Phi/2)$. } \cite{tHooft:1981bkw, tHooft:1979rtg}
\be
W(L)={1\over 2} Tr P \exp{(i g \oint_L A_k d x^k)} =  \cos{\Big({1\over 2} \ g \ \Phi\Big)}
\ee
is equal to $\Phi={4\pi \over g}$ when a closed loop $L$ is surrounding any oriented magnetic flux tube of the square area ${2\over a}  { \pi \over b}$ in the $(x,y)$ plane of the solution  (\ref{magsheet}) (see Fig.\ref{fig1}).  It will be convenient to call the solutions (\ref{magsheet} ) and (\ref{generasol})  "superfluxons"  as an abbreviation of "superposition of fluxes".  

When $\theta(X)=\arcsin({1\over  \cosh(a \cdot x)})$, we will obtain "hyperbolic" solution, which has infinite width in the $x$ direction compared with the finite width solution (\ref{polsol})
\be\label{hypersol}
n^a(x)= \{ {\cos((b\cdot x) \cosh^2(a\cdot x)) \over  \cosh(a\cdot x)}, { \sin((b\cdot x) \cosh^2(a\cdot x)) \over \cosh(a\cdot x)}, \tanh(a\cdot x)  \}.
\ee 
Finally,  when $\theta(X)= (a\cdot   x) $, we will obtain a  "trigonometric" solution 
\be\label{ansatz2}
n^a(\vec{x})= \{\sin(a\cdot x) \cos\Big({(b\cdot x) \over \sin(a\cdot x)}\Big),~\sin(a\cdot x) \sin\Big({(b\cdot x) \over \sin(a\cdot x)}\Big),~ \cos(a\cdot x)   \}. 
\ee 

\section{\it Topological properties of the solution }

Let us consider the topological properties of the solution  (\ref{generasol}), (\ref{magneticsheetsolution1}).  The conserved  topological  current and the corresponding magnetic charge can be defined in terms of the Abelian field strength $G_{\mu\nu}$ (\ref{chofact})  in analogy with the definition (\ref{topdensity})
 \beqa\label{chocurent}
K_{\mu} = {1 \over 2} \epsilon_{\mu\nu\lambda\rho} \partial_{\nu} G_{\lambda\rho}   =  {1 \over 2 g} \epsilon_{\mu\nu\lambda\rho} \partial_{\nu} S_{\lambda\rho}  
,  ~~~~~~\partial_{\mu}K_{\mu}=0,~~~~~~ Q_m= \int_{V} K_{0}d^3 x. 
 \eeqa
Here and in the next section we are considering the case of  vanishing Abelian field $F_{\mu\nu}=0$. In terms of the tensor $S_{\mu\nu}$ (\ref{spacetimefields}) and of the colour unit vector  $n^a$  (\ref{generasol}) the  topological charge will take the following equivalent forms:   
\beqa\label{topcharge}
&K_0={1 \over 2 g} \epsilon_{ijk} \partial_{i} S_{jk} = {1\over 2 g} \epsilon_{ijk} \partial_i (\epsilon^{abc}  n^a \partial_j n^b\partial_k n^c),\\
&Q_m=   {1\over 2 g} \int_{V} \epsilon_{ijk} \epsilon^{abc} \partial_i n^a \partial_j n^b\partial_k n^c d^3 x = {1\over 2 g} \int_{\partial V} \epsilon_{ijk} \epsilon^{abc}   n^a \partial_j n^b\partial_k n^c d \sigma_i= {1\over 2 g} \int_{\partial V}  d \sigma_i ~\epsilon_{ijk} S_{jk} .\nn
\eeqa
\begin{figure}
 \centering
\includegraphics[angle=0,width=12cm]{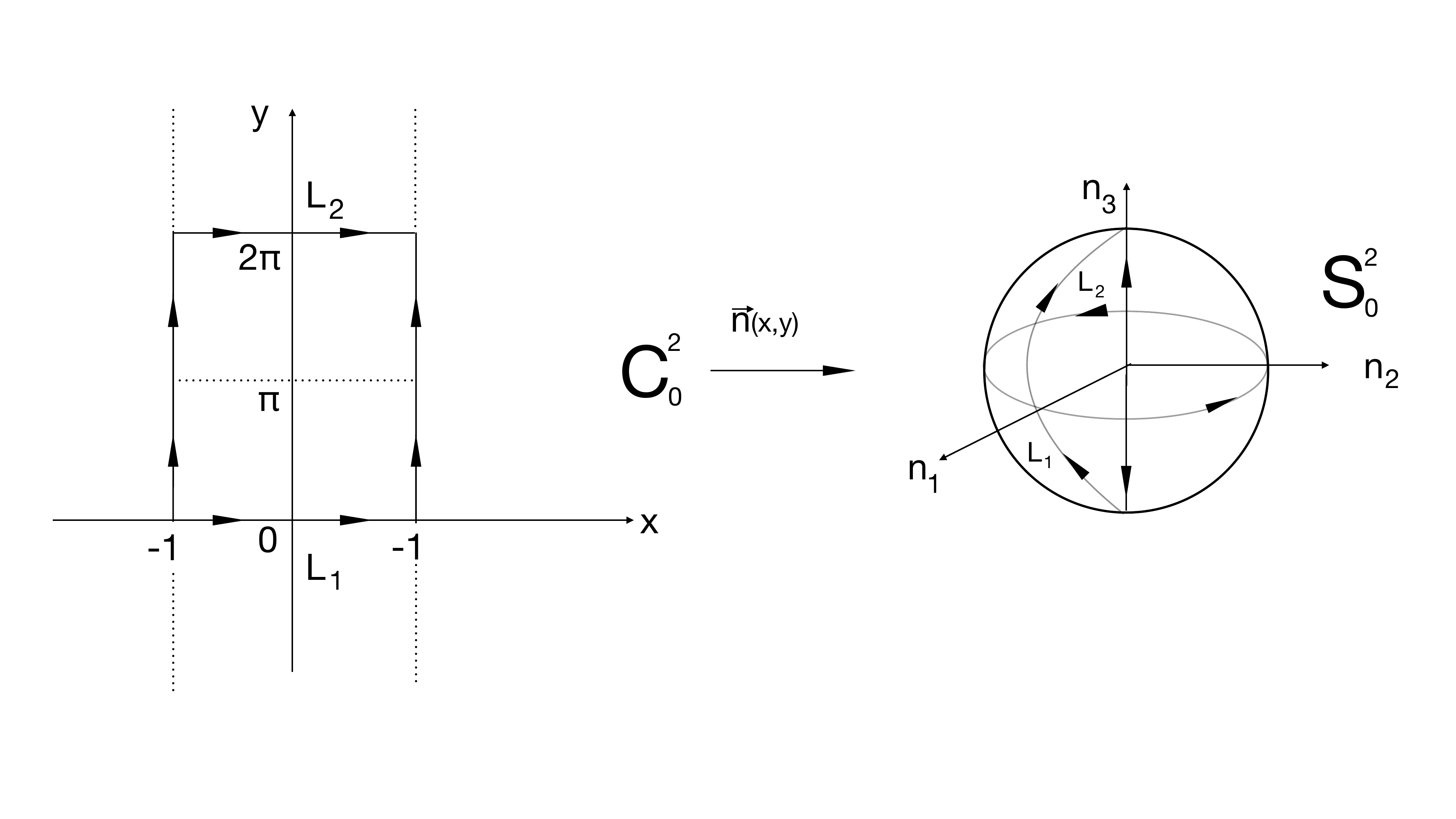}~~~~~~
\centering
\caption{The figure shows mapping  (\ref{polsol}) $ n^a(x,y)= \{ \sqrt{1-x^2} \cos(y),~ \sqrt{1-x^2} \sin(y),~x  \}$ of a cylinder cell $C^2_0$ on the plane $(x,y)$ to the sphere $S^2_0$. The mapping of the cylinder boundaries $x=\pm1$ to the north and south poles is given by the formula $n^a(\pm 1,y)=(0,0,\pm 1)$, where $y \in [0,2\pi]$.  The lines $L_{1,2}$ are identified on a sphere due to the formula $n^a(x,0)=n^a(x, 2\pi)= (\sqrt{1- x^2},0,x)$, where $x\in [-1,1]$.  Each cell $C^2_k$ defines magnetic fluxes in the $z$ direction (\ref{magfluxpos}).} 
\label{fig2} 
\end{figure}
 As far as the solution is homogeneous in the $z$ direction, we have to consider a magnetic flux through the space volume $V$  that is a rectangular box with its two boundaries being parallel to the $(x,y)$ plane at the distance $L$ from each other, and the other four boundaries will be defined for each particular solution individually. 
 
 Let us first consider the magnetic sheet solution (\ref{polsol}), (\ref{magsheet}). The rectangular boxes in this case will have four boundaries given by the equations $x= \pm{1\over a}$ and $y= [{2\pi  \over b}k, {2\pi  \over b}(k+1)]$, $k=0,\pm1, \pm2,... $ Because the tensor $S_{ij}$ is a space constant, the total charge $ Q_m=0$ (\ref{topcharge}).  For the solution (\ref{polsol}) a nonzero component of the tensor $S_{ij}$  is $S_{12}=-a b$ and $Q_m$ gets contributions only from two boundaries parallel to the $(x,y)$ plane:
 \be
 Q_m=    {1\over  g} \int_{\partial V}  S_{12} d \sigma_3 ={1\over  g} \int_{(x,y,0)}  a b~ dx dy -{1\over  g}  \int_{(x,y,L)}  a b~ dx dy= q_m(0) - q_m(L)=0.
 \ee
Thus  $q_m=q_m(0) = q_m(L)$, and we can define the invariant magnetic flux in terms of the surface integral:
\be
q_m ={1\over  g} \int_{(x,y,0)}  a b~ dx dy.
\ee
The vector $n^a(x,y)$ in (\ref{polsol}) defines a mapping of the cylinders $C^2_k: (x \times y) \in [-{1\over a},{1\over a}] \times [{2\pi \over b} k, {2\pi  \over b}(k+1)]$, $k=0,\pm1, \pm2,...$  into the spheres $S^2_k$, as it is shown in Fig.\ref{fig2}. After integrating over a given cylinder  $C^2_k$ of the area $\triangle x \triangle y =  (4 \pi /a b)$ one can see that the mapping  $n^a(x,y)$  covers  each sphere $S^2_k$ only once,  and the associated magnetic charge of this vortex in $z$ direction is
\be\label{magfluxpos}
g_m(k) = {1 \over  g} \int^{{1\over a}}_{-{1\over a}  }   ~d  a x    \int^{{2\pi \over b} (k+1)}_{{2\pi \over b} k}d b y={4\pi \over  g}. 
\ee
All magnetic charges $g_m(k)$ have the same sign.  Considering the alternative solution 
\be\label{antipolsol}
n^a(x)= \{ \sqrt{1-(a\cdot x)^2} \cos(b\cdot x),~ -\sqrt{1-(a\cdot x)^2} \sin(b\cdot x),~ (a\cdot x)  \},
\ee 
one can get convinced that this solution has the opposite magnetic charges:
\be
g_m(k) = - {4\pi \over  g}.
\ee
\begin{figure}
 \centering
\includegraphics[angle=0,width=9cm]{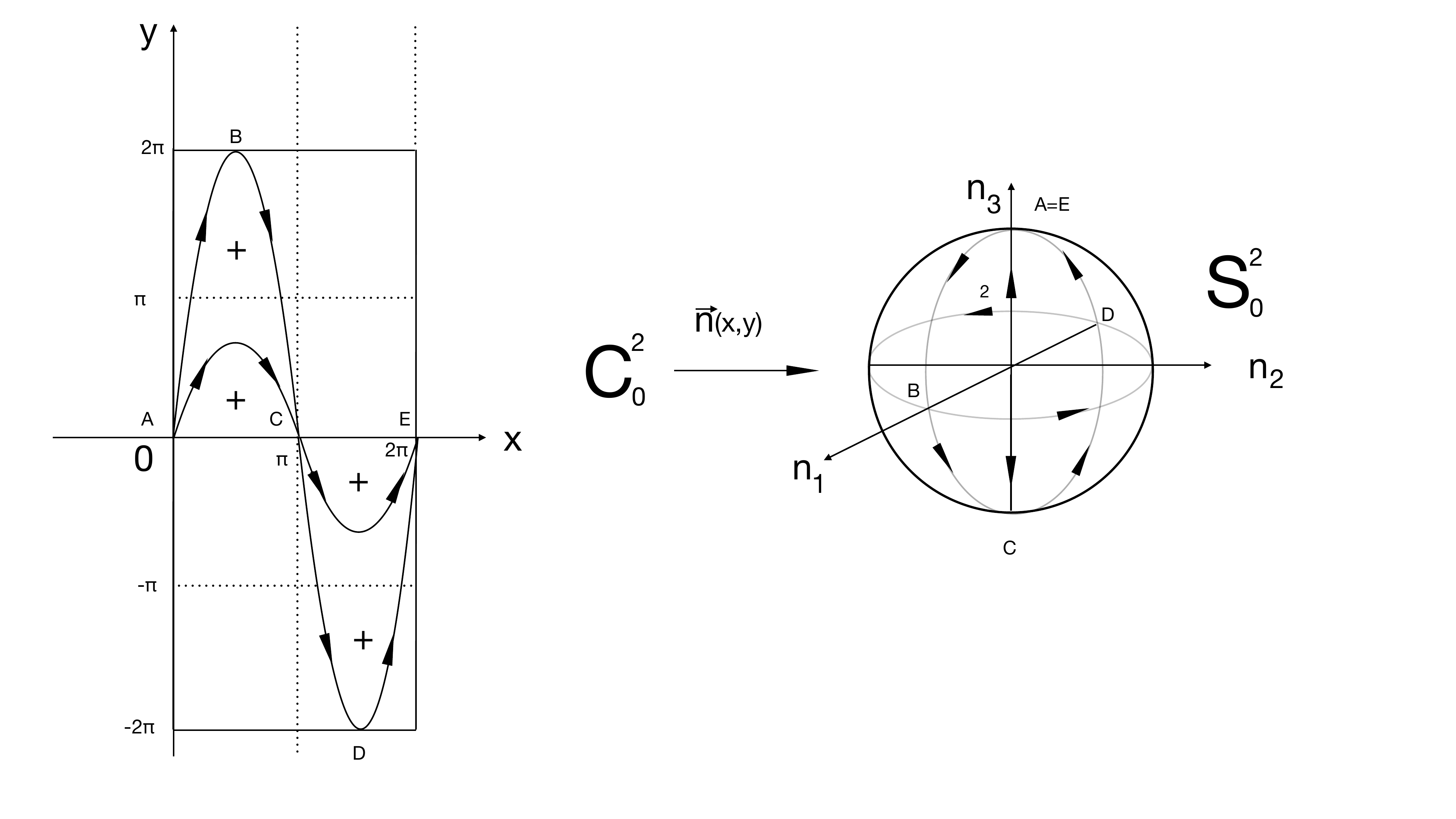}~
\includegraphics[angle=0,width=9cm]{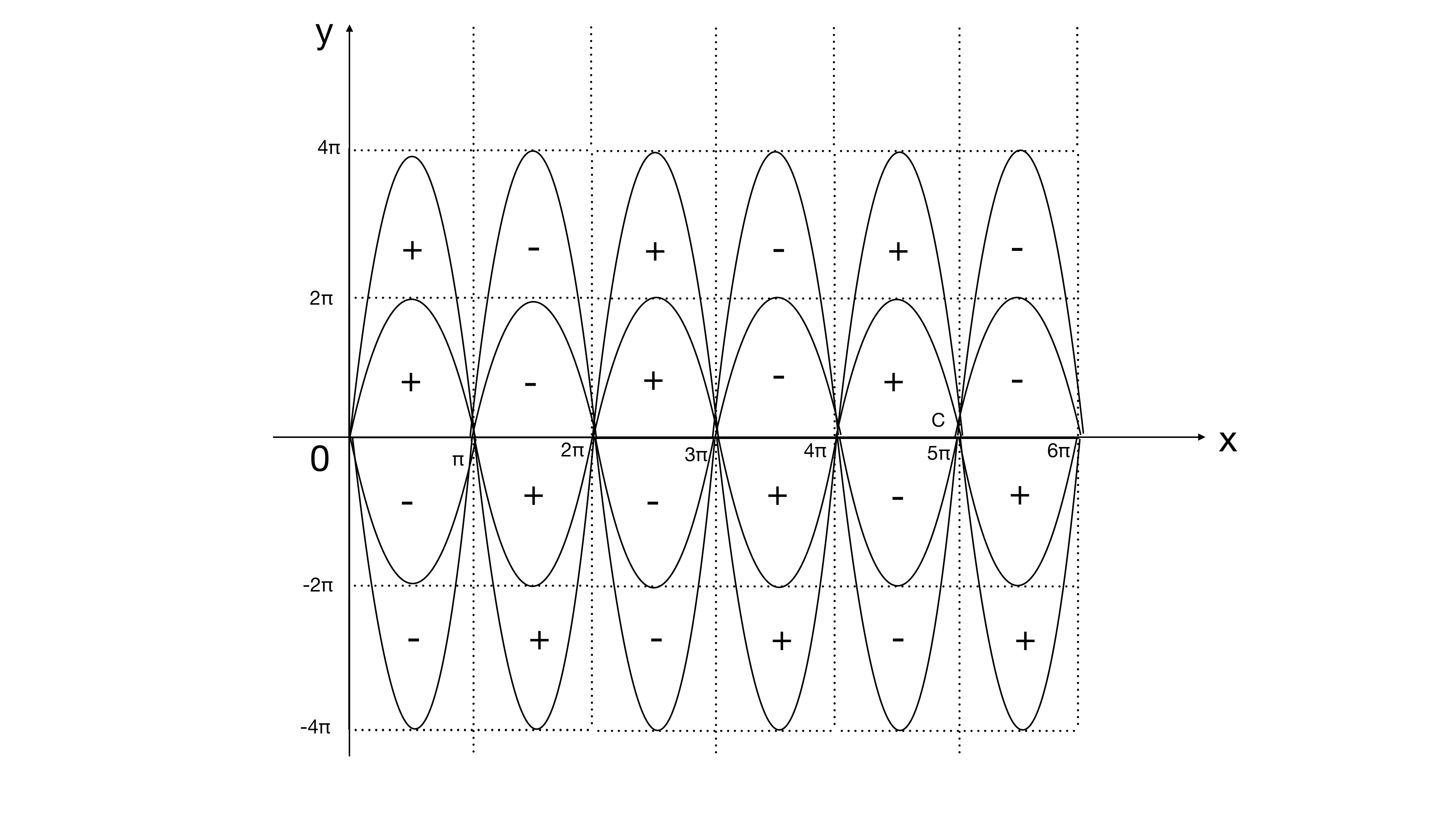}
\centering
\caption{The l.h.s figure shows the mapping defined by the vector $ n^a(x,y)= \{ \sin x \cos(y/\sin x),~ \sin x \sin(y/\sin x),~\cos x  \}$  from the cylinder cells $C^2_k$ to the spheres  $S^2_k$. The boundaries of the cylinders are defined by the equation $y=0, \pm \alpha \sin x$. The positive topological charges have the mapping of the cylinders  $y= 0,\alpha \sin x$ , $\alpha \in [0,2\pi k]$: ($n(0,\alpha) = (0,0,1)$,  $n(\pi,\alpha) = (0,0,-1)$, $n(\pi/2,\alpha) = (\cos \alpha,\sin \alpha,0)$, $n(3 \pi/2,\alpha) = (-\cos \alpha,-\sin \alpha,0)$, $n(2\pi,\alpha) = (0,0,1)$).  The negative topological charges have the mapping of the cylinders  $y= 0,-\alpha \sin x$, $\alpha \in [0,2\pi k]$).  The part of the full structure is shown on the r.h.s of the figure and reminds  the Abrikosov lattice of parallel Nielsen-Olesen magnetic vortices that are normal to the plane $(x,y)$ and have alternating magnetic fluxes.   }
\label{fig4} 
\end{figure}
Turning to the trigonometric solution (\ref{ansatz2}) one can find that here the cylinders have  the structure of $C^2_k: (x \times y) \in ([{2\pi \over a} k, {2\pi  \over a}(k+1)] \times [0, \pm \alpha \sin(a x)/b])$,  where $k=0,\pm1,...$, $\alpha \in [0,2\pi l]$, $l=0,\pm1, \pm2,...$,   as it is shown in Fig. \ref{fig4}. The important new property of this solution is that the magnetic fluxes have alternating magnetic charges. A similar pattern takes place for the general solution (\ref{generasol}), where the corresponding flux cylinders are defined by the equations  $f(a x)=(0, 2\pi k)$ and  $y=(0,  2 \pi l  f^{'}(ax) \sin f(ax) /b)$.  The magnetic flux tubes  are parallel  to each other, and the chromomagnetic fields inside the neighbouring vortices  are oriented in the opposite directions. In a sense, the solution describes a condensate of superposed Nielsen-Olesen magnetic vortices \cite{Nielsen:1973cs} of opposite magnetic fluxes Fig. \ref{fig2},\ref{fig4} and is a dual analog of the Cooper pairs condensate in a superconductor\footnote{ The existence of magnetic vortices in pure YM theory was predicted by using topological arguments based on the $Z_N$ twisted boundary conditions for the $SU(N)$ group in a finite box \cite{ tHooft:1979rtg}, as well as by analysing singularities in the so called Abelian projection gauge  \cite{  tHooft:1981bkw}.  The explicit solutions were suggested in \cite{tHooft:1981nnx}. }. 

\section{\it Potential barriers between solutions of constant energy density }

We have a large class of gauge field configurations that are defined by the function $\theta(X) $ and have a constant energy density  (\ref{energydenscont}).  We shall demonstrate that they are separated by potential barriers forming a complicated landscape.    In order to investigate this potential landscape we  will  perform a gauge transformation $U(\vec{x})$ that transforms the unit colour vector $n^a(x)$ (\ref{generasol}) into the constant vector in the third direction $n^{'a} = (0,0,1)$:
\be
\hat{n}^{'} = U^- \hat{n} U,~~~~~~~\hat{n} = n^a  \sigma^a.
\ee
The $SU(2)$  matrix of the corresponding singular gauge transformation has the following form:
\beqa\label{singtruns}
U=\left(
\begin{array}{cc}
\alpha&\beta\\\gamma&\delta
\end{array}
\right) = \left(
\begin{array}{cc}
\cos({f \over 2}) e^{{i\over 2} ({\pi \over 2} - {b y \over  f^{'}\sin f})}&i \sin({f \over 2}) e^{{i\over 2} ({\pi \over 2} - {b y \over f^{'}\sin f})}\\
i \sin({f \over 2}) e^{-{i\over 2} ({\pi \over 2} - {b y \over f^{'} \sin f})}&\cos({f \over 2}) e^{-{i\over 2} ({\pi \over 2} - {b y \over f^{'} \sin f})}
\end{array}
\right).
\eeqa
Under this gauge transformation $A^{'}_{\mu}= U^- A_{\mu} U - {i\over g} U^- \partial_{\mu} U$  the superfluxon gauge potential (\ref{magneticsheetsolution1})  will transform into the following form:
\beqa\label{magneticsheetsolution2}
A^{'3}_{1} =   {a b y \over g}   \Big({g H \over a b}  +\cot^2(f)    +     \cot(f)  {  f^{''}  \over f^{'2}  } \Big),
A^{'3}_{2} =   -{-b\over g}   {  \cot(f) \over f^{'}},~G^{'3}_{12} ={ ab- g H  \over g},\epsilon =   {(g H- a b)^2 \over 2 g^2}.~~~
\eeqa 
The nonzero components of the gauge field are now in the third colour direction $A^{'3}_{\mu}$. The Abelian gauge transformation $A^{''3}_{\mu} \rightarrow A^{'3}_{\mu}  - \partial_{\mu} \Lambda^3$,  where $\Lambda^3 = -b y{ \cot f\over f^{'}} $, brings the gauge potential (\ref{magneticsheetsolution2}) into a constant Abelian chromomagnetic field of the form (\ref{consfield}):
\be\label{magneticsheetsolution3}
A^{''3}_{1}=   {gH -a b\over g} y, ~~~A^{'3}_{2} = 0,~~~~~G^{'' 3}_{12} ={ ab- g H  \over g},~~~~~~\epsilon =   {(g H- a b)^2 \over 2 g^2}.
\ee  
The gauge configurations (\ref{magneticsheetsolution1}), (\ref{consfielstr}) and  (\ref{magneticsheetsolution2}), (\ref{magneticsheetsolution3}) are connected by a singular gauge transformation (\ref{singtruns}) and have the same chromomagnetic energy densities (\ref{energydenscont})  and (\ref{magneticsheetsolution2}), (\ref{magneticsheetsolution3}).  The question is: Should these field configurations be counted as physically identical or distinguishable in the functional integral over gauge field configurations?  The gauge-fixing procedure removes from the functional integral the gauge field configurations which can be joined by a continuous gauge transformation. But if the gauge fields  cannot be obtained from each other by gauge transformations which can be continuously joined by the identity transformation, then these gauge field configurations, although gauge equivalent,  should not be removed from the integrations over the field configurations by the gauge fixing procedure, as it was discussed by Jackiw and Rebbi \cite{Jackiw:1976pf, Callan:1976je}. In particular, the topologically distinguishable Chern-Pontryagin configurations are divided into several topologically inequivalent sectors separated by  potential barriers \cite{Jackiw:1976pf, Callan:1976je, Jackiw:1979ur}. 

\begin{figure}
 \centering
\includegraphics[angle=0,width=7cm]{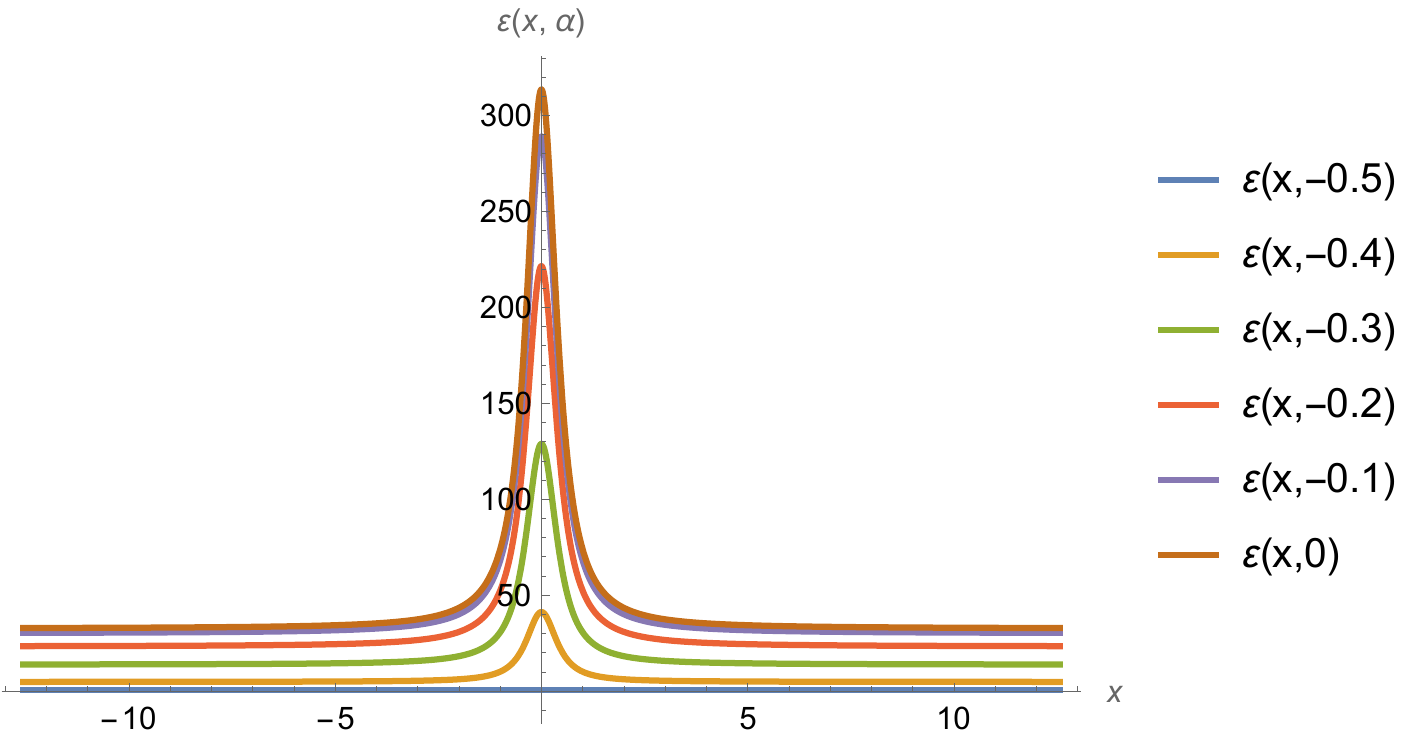}~~~~~~
\includegraphics[angle=0,width=7cm]{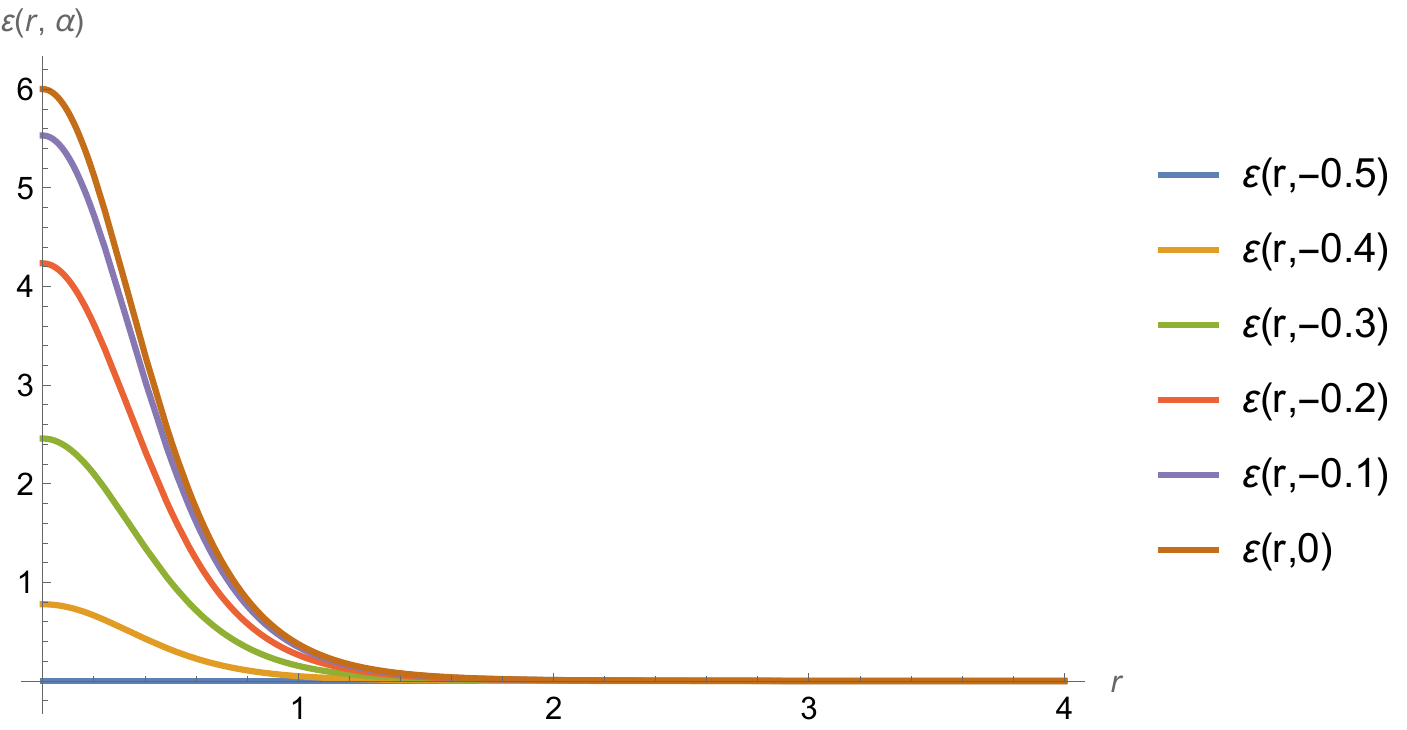}
\centering
\caption{The l.h.s. graph shows the shape of the  barrier $\epsilon(x,\alpha)$ (\ref{barriershape}) when $\alpha$  parameter  changes in the interval $[-{1\over 2}, 0]$. At  $\alpha=-{1\over 2}$ the energy density (\ref{energydenscont}) is  equal to $\epsilon   = 1/2$  ($a=b=g=1$). As $\alpha$  increases, the hight of the barrier increases and reaches its maximum at $\alpha=0$,  then it symmetrically decreases until $\alpha= {1\over 2}$, where it again is equal to $\epsilon   =   1/2$.   The r.h.s graph shows the shape of the  potential barrier (\ref{instbarrier}) between the Chern-Pontryagin vacua (\ref{flatcon}). 
}
\label{fig3} 
\end{figure}

The covariantly constant gauge field configurations are also divided into several topologically inequivalent sectors  and are separated by potential barriers.  Let us consider an arbitrary path $w(\alpha)$ that joins the field configurations (\ref{magneticsheetsolution1}) and (\ref{magneticsheetsolution2})  and calculate the corresponding magnetic energy density. To exemplify this, let us multiply the potential of  (\ref{magneticsheetsolution1}) by the  factor $w({1\over 2}- \alpha)$ and the potential (\ref{magneticsheetsolution2}) by the factor $ w({1\over 2}+ \alpha)$  requiring  that the $w(0)=0$ and $w(1)=1$, when $\alpha$ increases   from $-{1\over 2}$ to $+{1\over 2}$:
\beqa\label{pathdeformation}\
\hat{A}^{a}_{\mu}= w({1\over 2}- \alpha) A^{a}_{\mu}+w({1\over 2}+ \alpha) A^{'a}_{\mu}.
\eeqa 
These factors define a path  $w(\alpha)$ that connects field configurations (\ref{magneticsheetsolution1}) and (\ref{magneticsheetsolution2}) and allows to investigate the energy landscape of covariantly  constant gauge fields configurations by calculating the magnetic energy along this path,  when $\alpha =-{1\over 2}$ the $\hat{A}^{a}_{\mu}$ coincides with $A^{a}_{\mu}$ and when $\alpha ={1\over 2}$ it coincides with $A^{'a}_{\mu}$. After substituting the field $(\ref{pathdeformation})$ into the energy density functional $\epsilon =  {1\over 4 }G^{a}_{ij} G^{a}_{ij}$ we will get the following shape of the potential barrier\footnote{ We present a compact expression when the Abelian component $F_{\mu\nu}=0$, otherwise the expression is larger. }:
\beqa\label{genebarriershape}
\epsilon(x,\alpha)={a^2 b^2 \over 2 g^2} \Big( (2 -w_{-})^2 w^2_{-}  +  
 2 (2 -w_{-})w_{-} (1 +w_{-} )w_{+}  \cos f+ (1+ w^2_{-} \cot^2f) w^2_{+}\Big),~~
\eeqa
where $w_- \equiv  w({1\over 2}- \alpha)$ and   $w_+ \equiv w({1\over 2}+ \alpha)$.  The barrier is homogeneous in $y$ and $z$ directions. The  profile of the potential barrier between superfluson (\ref{magneticsheetsolution1}) at $H=0$ and the flat configuration (\ref{magneticsheetsolution2}) depend on the behaviour of $ \cos f(ax)$ and ${1\over \sin^2 f(ax)} $ and of the moduli parameter $f(ax)$. It follows that $\epsilon(x,\pm 1/2)= {a^2 b^2 \over 2 g^2}$   in accordance with the energy densities (\ref{energydenscont})) at $H=0$ and (\ref{magneticsheetsolution2}).    If $w(\alpha)$ is a linear functional of its argument $w(\alpha) =  \alpha$, then we will get\footnote{There is no potential barrier between Abelian fields $A^{'a}_{\mu}$ and $A^{''a}_{\mu}$. The topological structure of the field strength tensors and their singularities  define the structure of the potential barriers.  The field strength tensors $G^{' 3}_{12}$ (\ref{magneticsheetsolution2})  and $G^{" 3}_{12}$ (\ref{magneticsheetsolution3}) of the solutions $A^{'}_{\mu}$ and $A^{"}_{\mu}$ are in the same topological class, because the gauge transformation is trivial between them $U=1$.  The gauge transformation between the field strength tensors (\ref{consfielstr}) and (\ref{magneticsheetsolution2}), (\ref{magneticsheetsolution3}) is topologically nontrivial and   singular (\ref{singtruns}). As a consequence,  the corresponding configurations are separated by potential barrier (\ref{genebarriershape}). } 
\be\label{barriershape}
\epsilon(x,\alpha)= {a^2 b^2 \over 32 g^2} \Big( 12 - 8 \alpha + 16 \alpha^2 + 36 \alpha^3  +  (18 - 80 \alpha^2 + 32 \alpha^4) \cos f(ax) + {(1-4 \alpha^2)^2 \over \sin^2 f(ax)}~ \Big). 
\ee
For a particular solution $f(ax)$ the potential barrier is shown in Fig.\ref{fig3}. 

\section{\it Potential barriers between vacuum solutions  }

New phenomena appear in the case of non-vanishing   Abelian field   $B_{i}=-{1\over 2} F_{ij} x_{j} $ in (\ref{choansatz}),  (\ref{spacetimefields}). The  solutions (\ref{choansatz}),  (\ref{consfield}), (\ref{generasol}) and (\ref{magneticsheetsolution1}) can be considered as  an exact non-perturbative solution of the Yang Mills  equation in the background chromomagnetic field \cite{Savvidy:2024sv, Savvidy:2024ppd}.    The  solution is given by the sum (\ref{choansatz})  $A^a_i = -{1\over 2} F_{ij} x_{j} n^a + {1\over g} \varepsilon^{abc} n^{b} \partial_{i}n^{c} $, $A^a_0 =0$, where $n^a$ is defined by the equation (\ref{generasol}).  The solution is parametrised by three vectors $\vec{H}$, $\vec{a}$ and $\vec{b}$, and the magnetic energy density has the   general form (\ref{genenergden}) :
\beqa\label{trigenergydens}
\epsilon =  {1\over 4 }G^{a}_{ij} G^{a}_{ij}  &=&   {(g  \vec{H} -   \vec{a} \times \vec{b} )^2 \over 2 g^2} .
\eeqa  
This means that the  magnetic energy  density ${1\over 2}H^2$  of the Abelian field is lowered by the nonlinear interaction and the zero energy density $\epsilon$ is realised when 
\be\label{vacuumcond}
g  \vec{H}_{vac} =  \vec{a}\times \vec{b}, ~~~~~\epsilon({g  \vec{H}_{vac} })= 0.
\ee
This takes place when three vectors  $ (\vec{H}, \vec{a}, \vec{b})$ are forming an orthogonal right-oriented frame.  {\it  At the minimum (\ref{vacuumcond}) the field strength tensor vanishes, $G_{ij}=0$, and  the general solution (\ref{choansatz}),  (\ref{consfield}), (\ref{generasol}),  (\ref{magneticsheetsolution1}) reduces to a  flat  vacuum connection of the following form}: 
\beqa\label{flatconnection} 
A^{a}_{i} &=&   {1\over g} \left\{
\begin{array}{cccc}   
 \Big(a b y \csc f \cos \left(\frac{b y \csc f}{f'}\right) - a f' \sin \left(\frac{b y \csc f }{f'}\right)+ \frac{a b y f''  }{f'^2}  \cos f \cos \left(\frac{b y \csc f}{f'}\right) ,  \\ 
 a b y \csc f \sin \left(\frac{b y \csc f}{f'}\right) + a f' \cos \left(\frac{b y \csc f }{f'}\right)+ \frac{a b y f'' }{f'^2}  \cos f \sin \left(\frac{b y \csc f}{f'}\right), 
  -a b y  \frac{f''  \sin f }{f'^2} \Big)  ~~~~ \\
{1\over f'}\Big(-b  \cos f  \cos \left(\frac{b y \csc f }{f' }\right),-b \cos f \sin \left(\frac{b  y \csc f }{f' }\right),   b\sin f  \Big)&\\
(0,0,0)&
\end{array}, \right. 
\eeqa 
where $a$ and $b$ are the scale parameters of the moduli space and are similar to the moduli parameters of the instanton solution.  The flat connection (\ref{flatconnection}) can be represented in the standard form 
\be\label{flatconnection1}
\vec{A}_{g\vec{H}= \vec{a}\times \vec{b}} ~=~ -{i \over g}  U^{-}  \vec{\nabla} U
\ee
 and is characterised by the vector $g\vec{H}_{vac}= \vec{a}\times \vec{b}$ in the 3d-space. 
Away from the vacuum flat connection (\ref{flatconnection} ) the energy  density increases quadratically (\ref{trigenergydens}).  Here as well one can investigate the details of the potential barriers between these vacuum solutions. When the Abelian part $B_{1}=H y$  of the gauge potential  (\ref{choansatz}) is present,  then the potential barrier will take the following form:
\beqa\label{barriershapeH}
 \epsilon(g H,x,y,\alpha)&=& {a^2 b^2 \over 32 g^2} \Big( (12 - 8 \alpha + 16 \alpha^2 + 32 \alpha^3 - 8(1+4 \alpha^2){g H \over a b}~)(1- {g H \over a b}) +\\
&+& 2  (1-4 \alpha^2) ((2{g H \over a b} -3 )^2 - 4 \alpha^2) \cos f(ax) + {(1-4 \alpha^2)^2 \over \sin^2 f(ax)} + {(1-4 \alpha^2)^2  g^2 H^2 y^2\over a^2  f^{'}_x(ax)^2} \Big). \nn
\eeqa
At $H=0$ it reduces to the previous expression  (\ref{barriershape}) $\epsilon(0,x,y,\alpha)= \epsilon(x,\alpha )$ and   at  $\alpha = \pm 1/2$ we have 
\be\label{min1}
\epsilon(g H,x,y, \pm 1/2) = {( g H -  a b )^2\over 2 g^2} . 
\ee 
At $g H  =  a b$  the initial  and final configurations  (\ref{flatconnection}) and  (\ref{magneticsheetsolution2})  are vacuum flat connections because for them $\epsilon =0$ and the energy barrier between these vacuum configurations  is
\beqa\label{barriershapeH}
 \epsilon(gH=a b,x,y,\alpha)&=&  
  (1-4 \alpha^2)^2  {a^2 b^2 \over 32 g^2}    \Big( 2  \cos f(ax) + {1\over \sin^2 f(ax)} + {    b^2   y^2\over   f^{'}_x(ax)^2} \Big). 
\eeqa
One can conjecture that due to the tunnelling transitions between degenerate   vacua (\ref{flatconnection}), which have different orientations of the vector $g\vec{H}= \vec{a}\times \vec{b}$, a new  quantum state  will be induced
$
\psi(\vec{A}) = \int \CG_{  \{ \phi,\theta,\chi \} }~ \psi  (A_{g\vec{H}= \vec{a}\times \vec{b}}) ~d \mu( \phi,\theta,\chi),
$
where $\CG_{  \{ \phi,\theta,\chi \} }$ is the space rotations operator.  It is  a quantum-mechanical superposition of the states  $\psi  (A_{g\vec{H}= \vec{a}\times \vec{b}})$ with different orientations of the vector $g\vec{H}= \vec{a}\times \vec{b}$.  The Lorentz invariance of the vacuum state would be restored  at the quantum-mechanical level. 

The CP violating  topological effect appeared due to the presence of vacuum  gauge field configurations that have non-vanishing  Chern-Pontryagin  index   \cite{tHooft:1976rip, Jackiw:1976pf, Callan:1976je, Jackiw:1979ur}:  
\be\label{flatcon} 
\vec{A}_n(\vec{x}) =-{i \over g} U^{-}_n(\vec{x}) \nabla U_n(\vec{x}), ~~~~ U_1(\vec{x})= {\vec{x}^2 -\lambda^2  - 2 i \lambda \vec{\sigma} \vec{x} \over \vec{x}^2 +\lambda^2}, ~~~~~U_n = U^n_1.
\ee
The values of the gauge field  (\ref{flatcon}), although gauge equivalent to $\vec{A}(x) = 0$, are not removed from the integration over the field configurations by gauge fixing procedure because they belong to different topological classes and are separated by potential barriers \cite{tHooft:1976rip, Jackiw:1976pf, Callan:1976je, Jackiw:1979ur}.  The potential barriers can be  calculated between these field configurations considering $\vec{A}^{~'}_1(\vec{x}) = ({1\over 2} - \alpha)\vec{A}_1(\vec{x})$  when $\alpha$ is continuously varying from $-{1\over2}$ to ${1\over2}$. The potential barrier has the following shape  
\be\label{instbarrier}
\epsilon(r, \alpha) = {1\over 4} G^a_{ij}G^a_{ij}= {6 \lambda^4   (1- 4 \alpha^2)\over g^2 (r^2 +\lambda^2)^4 }
\ee
shown in Fig.\ref{fig3}. In the quantum theory  tunnelling will occur across this barrier and the  quantum-mechanical superposition $\Psi_{\theta}(\vec{A}) = \sum_n e^{i n \theta} \psi_n(\vec{A})$ represents the  Yang Mills $\theta$ vacuum state  \cite{Jackiw:1976pf, Callan:1976je}. The induced Chern-Pontryagin $\theta$-angle term is Lorentz invariant, but breaks the CP invariants, so that the distinct $\theta$ vacuum states correspond to distinct theories \cite{tHooft:1976rip, Jackiw:1976pf, Callan:1976je, Jackiw:1979ur}. 

 We do not know yet whether there exist  the instanton-like transitions that would induce a tunnelling between   vacuum configurations with nonzero Pontryagin  index  (\ref{flatcon}) and the "superfluxon" vacuum configurations (\ref{flatconnection}),  (\ref{flatconnection1}).  A possible tunnelling  transition between superfluxon  flat configurations and the flat configurations with non-vanishing  Chern-Pontryagin index (\ref{flatcon} ) will wash out the CP violating $\theta$ angle to zero, dynamically restoring CP symmetry.  

\section{\it Vacuum polarisation}

The existence of an even larger class of  covariantly constant gauge field configurations pointed out to the fact that the Yang-Mills vacuum has even  larger  degeneracy of vacuum field configurations\footnote{The Ising spin system that has {\it an exponential degeneracy of its vacuum configurations}  was discovered  in \cite{Savvidy:1993ej}.  Here the parallel planes of differently oriented spin configurations  represent the degenerate vacuum spin configurations which are separated by potential barriers  \cite{Savvidy:2000zq}.  The total number of such vacuum configurations is $3 \times 2^{N}$ or  $ 2^{3 N}$ if $k=0$   \cite{Savvidy:2000zq, Savvidy:1993sr, Savvidy:1994sc,Savvidy:1994tf, Pietig:1996xj, Pietig:1997va}.   In recent publications this symmetry was referred to as the {\it subsystem symmetry}, and it has exotic fracton excitations \cite{Sherrington:1975zz, 2010PhRvB..81r4303C, Vijay:2016phm}. }. It is a challenging problem to investigate  the vacuum polarisation induced by the new class of covariantly constant gauge fields.  The early investigation  revealed that the effective Lagrangian of the SU(N) Yang-Mills  theory  has the following gauge and Lorentz invariant form:
\beqa\label{YMeffective}
\CL  =  
-\CF - {11  N \over 96 \pi^2}  g^2 \CF \Big( \ln {2 g^2 \CF \over \mu^4}- 1\Big),
\eeqa
where  $\CF= {1\over 4} G^a_{\mu\nu}G^a_{\mu\nu} ={\vec{\CH}^2_a -\vec{\CE}^2_a\over 2}  \geq 0$ and   $\CG = {1\over 4} G^a_{\mu\nu} \tilde{G}^a_{\mu\nu} = \vec{\CH}_a \vec{\CE}_a =0$, and that the vacuum energy density has its new minimum  at a nonzero value of the field strength \cite{Savvidy:1977as}:
\be\label{chomomagneticcondensate}
 \langle 2  g^2 \CF \rangle_{vac}=    \mu^4  \exp{(-{96 \pi^2 \over 11 N g^2(\mu) })} = \Lambda^4_{S},~~~~\epsilon_{vac}= - {11 N \over 192 \pi^2}  \Lambda^4_{S}.
\ee
One can conjecture that the effective Lagrangian for the general covariantly constant  gauge fields has a universal form  (\ref{YMeffective}) dynamically  breaking  the conformal invariance of the solutions.  

In that respect it is important to stress that {\it the Yang Mills effective Lagrangian is gauge invariant only on the exact solutions of the quantum sourceless Yang Mills equation} \cite{Batalin:1976uv,Batalin:1979jh}. Therefore the exact solutions (\ref{magneticsheetsolution1}) can play a crucial role in the search of the Yang Mills theory vacuum state. The other class of solutions were considered in \cite{ Pak:2020izt, Pak:2020obo, Pak:2017skw, Pak:2020fkt,  Anous:2017mwr, Kim:2016xdn, Milshtein:1983th,   Apenko:1982tj, Reuter:1994yq, Reuter:1997gx,   Wu:1975vq, Wu:1967vp} and in \cite{Baseian:1979zx, Savvidy:1982wx, Savvidy:1982jk, Matinyan:1981ys, Matinyan:1981dj,Banks:1996vh}. 

The calculation of the effective Lagrangian by using the general ansatz (\ref{choansatz}) cannot be justified because the results will be not gauge invariant and therefore will be unphysical. The problem of gauge invariance of the effective Lagrangian in gauge field theory was intensively discussed in the literature \cite{Jackiw:1974cv, Drummond:1974sw, Nielsen:1975fs, Batalin:1976uv, Batalin:1979jh, Abbott:1980hw}.

In conclusion I would like to thank Konstantin Savvidy, Youngmin Cho, Mikhail Shaposhnikov and Alexander Migdal for stimulating discussions.  

\bibliographystyle{elsarticle-num}
\bibliography{magnetic_PLB}

\end{document}